\documentclass[reprint,prl,aps,superscriptaddress]{revtex4-1}
\usepackage{times}
\usepackage{graphicx}
\usepackage{amsmath, braket, amsfonts}
\usepackage{amssymb}
\usepackage{natbib}
\usepackage{sidecap}
\usepackage{bm, color}
\usepackage{xcolor}
\usepackage{ragged2e}
\usepackage{lipsum}
\usepackage{wrapfig,booktabs}
\usepackage{siunitx}
\usepackage{xspace}
\usepackage{placeins}
\newcommand{\abs}[1]{\left\lvert #1 \right\rvert}
\newcommand{\ed}{\ensuremath{\varepsilon_d}\xspace}
\newcommand{\eq}{\ensuremath{\varepsilon_m}\xspace}
\newcommand{\ics}{\ensuremath{I_{m}}\xspace}
\newcommand{\bpar}{B_\parallel}

\begin{document}

\title{Anderson Orthogonality as Measurement Backaction in Coupled Quantum Dots}

\author{Will Grant}
	\affiliation{Stewart Blusson Quantum Matter Institute, University of British Columbia, Vancouver, British Columbia, V6T1Z4, Canada}
	\affiliation{Department of Physics and Astronomy, University of British Columbia, Vancouver, British Columbia, V6T1Z1, Canada}
\author{Sarath Sankar}
	\affiliation{School of Physics and Astronomy, Tel Aviv University, Tel Aviv 6997801, Israel}
	\affiliation{Department of Physics, Ben-Gurion University of the Negev, Beer Sheva 84105, Israel}
\author{Elena Cornick}
	\affiliation{Stewart Blusson Quantum Matter Institute, University of British Columbia, Vancouver, British Columbia, V6T1Z4, Canada}
	\affiliation{Department of Physics and Astronomy, University of British Columbia, Vancouver, British Columbia, V6T1Z1, Canada}
\author{Vahid Movahed}
	\affiliation{Stewart Blusson Quantum Matter Institute, University of British Columbia, Vancouver, British Columbia, V6T1Z4, Canada}
	\affiliation{Department of Physics and Astronomy, University of British Columbia, Vancouver, British Columbia, V6T1Z1, Canada}
\author{Johann Drayne}
	\affiliation{Stewart Blusson Quantum Matter Institute, University of British Columbia, Vancouver, British Columbia, V6T1Z4, Canada}
	\affiliation{Department of Physics and Astronomy, University of British Columbia, Vancouver, British Columbia, V6T1Z1, Canada}
\author{Silvia L\"{u}scher}
	\affiliation{Stewart Blusson Quantum Matter Institute, University of British Columbia, Vancouver, British Columbia, V6T1Z4, Canada}
	\affiliation{Department of Physics and Astronomy, University of British Columbia, Vancouver, British Columbia, V6T1Z1, Canada}
\author{Saeed Fallahi}
	\affiliation{Department of Physics and Astronomy, Purdue University, West Lafayette, Indiana, USA}
\author{Geoffrey C. Gardner}
	\affiliation{Microsoft Quantum, West Lafayette, Indiana, USA}
\author{Michael J. Manfra}
	\affiliation{Department of Physics and Astronomy, Purdue University, West Lafayette, Indiana, USA}
     \affiliation{Microsoft Quantum, West Lafayette, Indiana, USA}
    \affiliation{Elmore Family School of Electrical and Computer Engineering, Purdue University, West Lafayette, Indiana, USA}
    \affiliation{School of Materials Engineering, Purdue University, West Lafayette, Indiana, USA}
    \affiliation{Purdue Quantum Science and Engineering Institute, Purdue University, West Lafayette, Indiana, USA}
\author{Eran Sela}
	\affiliation{School of Physics and Astronomy, Tel Aviv University, Tel Aviv 6997801, Israel}
\author{Yigal Meir}
	\affiliation{Department of Physics, Ben-Gurion University of the Negev, Beer Sheva 84105, Israel}
\author{Joshua Folk}
    \email{jfolk@physics.ubc.ca}
	\affiliation{Stewart Blusson Quantum Matter Institute, University of British Columbia, Vancouver, British Columbia, V6T1Z4, Canada}
	\affiliation{Department of Physics and Astronomy, University of British Columbia, Vancouver, British Columbia, V6T1Z1, Canada}

\date{\today}

\begin{abstract}

Measurement perturbs a quantum system by coupling it to external degrees of freedom, but detector backaction depends on the physical mechanism of measurement itself.  In solid-state devices, detectors driven far from equilibrium to enable faster measurements produce backaction that can often be understood as classical noise.  However, a strong measurement can also induce backaction from quantum many-body correlations in the detector that are intrinsic to the measurement, even without shot noise.  Here, we probe this near-equilibrium backaction through the effect of a quantum-dot charge sensor on tunnelling between a second quantum dot and its reservoirs.  The measurement realizes the Anderson Orthogonality Catastrophe (AOC): electrons in the detector leads reorganize in response to an abrupt change in local scattering potential, suppressing resonant tunnelling while enabling inelastic processes that exchange energy with the detector.  Changing the detector energy level tunes the AOC backaction from negligible to dominant in the tunnelling dynamics.  More broadly, these results establish detector-induced many-body correlations as a controllable influence on quantum dynamics.

\end{abstract}

\maketitle

Measuring a quantum system requires coupling it to external degrees of freedom.  The same coupling that makes information about the system available also modifies its otherwise unitary evolution, an effect known as measurement backaction.  Seminal demonstrations of this principle in atomic systems have revealed discrete quantum jumps associated with wavefunction collapse\cite{nagourney1986shelved,bergquist1986observation} and the suppression of coherent evolution by continuous observation–the quantum Zeno effect\cite{itano1990quantum}.  In solid-state quantum devices, measurement typically couples a coherent quantum system to a nearby circuit element whose electrical response depends on the charge state of the system.  When this circuit element is a conductor biased far from equilibrium, its backaction is typically associated with shot-noise-driven fluctuations that suppress coherence, and is often described as a source of classical noise~\cite{Aleiner1997Dephasing,Buks1998Dephasing,ji2003electronic,AvinunKalish2004Controlled,Kung2009Noise-induced,Bischoff2015Measurement,Ferguson2023Measurement-induced,Sankar2024Detector-tuned,Sankar2025Back-action}.

But detector noise is not the only way that measurement can disturb a quantum system: a conducting detector is a many-electron environment, and changing the charge of the measured system changes the scattering potential experienced by electrons in the detector.  The resulting collective rearrangement in the detector and its reservoirs is the Anderson Orthogonality Catastrophe: the detector states associated with different system charge configurations become nearly orthogonal in the many-body Hilbert space, providing a form of backaction intrinsic to the detection mechanism.\cite{Anderson1967Infrared,Nozieres1969,Aleiner1997Dephasing,Sankar2024Detector-tuned,Sankar2025Back-action,Sankar2025Direct}  Here we explore this intrinsic form of backaction in a mesoscopic quantum circuit using a quantum-dot detector held near equilibrium, where conventional noise backaction is strongly reduced.  By tuning the detector energy level, we modify how strongly the detector scattering phase shift changes when the measured charge changes, allowing Anderson-orthogonality backaction to be tuned from weak to strong limits and measured directly.

\begin{figure}[t!]
    \includegraphics[width=\columnwidth]{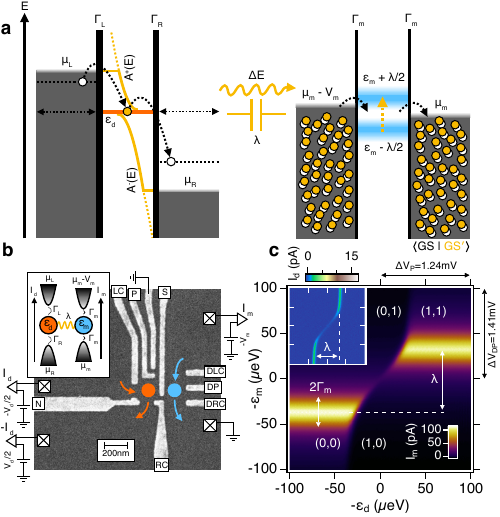}
    \caption{\textbf{Experimental overview.} \textbf{a}, Mechanism by which Anderson Orthogonality Catastrophe modifies tunnel rates. An electron tunnelling onto the dot (left) forces an abrupt restructuring of the detector Fermi sea (right) from one many-body ground state to another. The corresponding spectral functions, $A^{\pm}(E)$, describe energy exchange with  the detector during tunnel-in and tunnel-out processes, modifying the bare tunnel rates $\Gamma_{L,R}$. \textbf{b}, Scanning electron micrograph of the device and measurement circuit. The device schematic (inset) illustrates a dot (orange) weakly coupled to leads, and capacitively coupled to the detector (blue) strongly coupled to its own leads.
    \textbf{c}, Charge stability diagrams near equilibrium ($V_d=3 \mu$V, $V_m=5 \mu$V), showing simultaneous measurements of $I_{m}$ (main panel) and $I_{d}$ (inset) across the coupled $(N_{d},N_{m})$ charge transitions.}
    \label{fig:1}
\end{figure}

In our experiment (Fig.~\ref{fig:1}), the measured system is a quantum dot, weakly tunnel-coupled to its leads, that may contain either zero or one electron; we refer to this system dot as the ``dot" (subscript $d$). Its occupation, $N_d$, is monitored by a second quantum dot that is capacitively coupled to the first, strongly tunnel-coupled to its reservoirs, and held near equilibrium; we refer to this second dot as the ``detector” (subscript $m$)~\cite{Kung2009Noise-induced,Bischoff2015Measurement,Ferguson2023Measurement-induced}.  Adding or removing an electron from the dot abruptly shifts the detector energy level, changing the scattering phase shift of electrons in the detector reservoirs and forcing a collective restructuring of the detector Fermi sea.  This Anderson-orthogonality response suppresses resonant tunnelling through the dot while enabling inelastic processes in which energy is exchanged with the detector~\cite{Anderson1967Infrared,Nozieres1969,Sankar2025Direct}.

\begin{figure*}[t]
    \includegraphics[width=\textwidth]{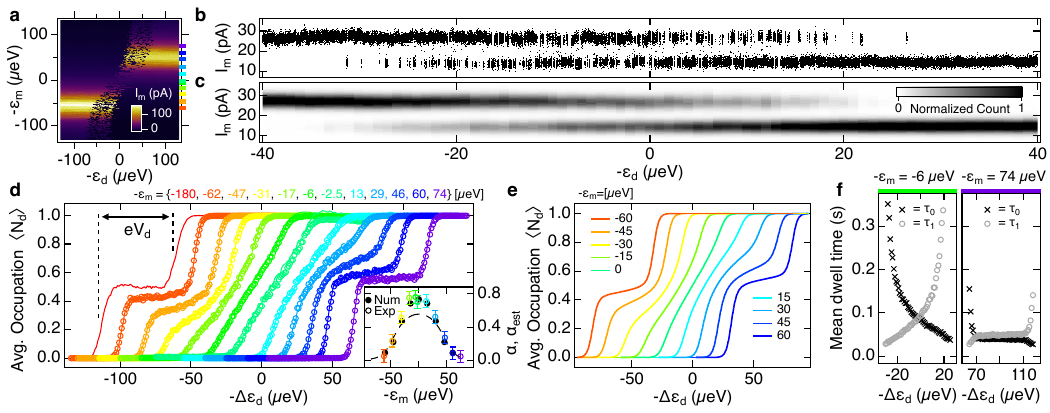}
    \caption{\textbf{Time-resolved measurements of AOC-modified tunnelling.} 
\textbf{a}, Detector current map measured with a voltage bias $V_d=50 \ \mu $V across the dot, for $\lambda=104 \ \mu e$V and $\Gamma_m=22 \ \mu e$V. Individual tunnelling events are resolved when \ed lies within the bias window. 
\textbf{b}, Detector current during a slow sweep of \ed across the charge transition ($-\eq=-6 \ \mu e$V), revealing individual hopping events. 
\textbf{c}, Histograms of detector current at each \ed, showing the bimodal distribution of $N_d=0$ and $N_d=1$ charge states ($-\eq=-6 \ \mu e$V). 
\textbf{d}, Average occupation $\langle N_d \rangle(\ed)$ for several detector detunings \eq, indicated by colour in panel \textbf{a}. Markers are extracted from the histograms; solid lines are obtained from normalized $\langle I_m \rangle$. Inset: analytic AOC exponent from Eq.~\ref{eq:AOC_delta} (dashed line), compared with susceptibility-based estimates from experiment and numerical calculations (markers). 
\textbf{e}, Calculated $\langle N_d \rangle(\ed)$ for comparable detector detunings, following Ref.~\cite{Sankar2025Direct}. 
\textbf{f}, Average dwell times in the $N_d=0$ and $N_d=1$ states versus \ed, for detector settings corresponding to strong and weak AOC. Coloured bar above each graph indicates the corresponding trace from panel \textbf{d}.}
    \label{fig:tr}
\end{figure*}

The central experimental signature is the transformation of tunnelling dynamics into and out of the dot as the detector is tuned from far from resonance to near resonance.  To reveal this transformation, we apply a voltage bias across the dot, exposing an energy window over which tunnelling can occur.   When the detector is far off resonance, tunnelling through the dot resembles ordinary resonant transport: the tunnel rates are nearly energy independent, and the dot occupation forms a plateau across this bias window.  When the detector is tuned near resonance, the detector Fermi sea is significantly rearranged by each change in dot charge, and the plateau is replaced by a strongly energy-dependent occupation.  This evolution provides a direct way to measure the many-body backaction induced by the detector.

This transformation is controlled by how strongly the dot charge modifies the scattering properties of the detector.  Adding an electron to the dot shifts the detector energy level by $\lambda$ through capacitive coupling.  The corresponding difference in detector scattering phase shift between the zero- and one-electron dot states is
\begin{equation}
    \delta = \tan^{-1}\!\left( \frac{\varepsilon_m + \lambda/2}{\Gamma_m} \right)
           - \tan^{-1}\!\left( \frac{\varepsilon_m - \lambda/2}{\Gamma_m} \right),
    \label{eq:AOC_delta}
\end{equation}
where $\Gamma_m$ is the broadening of the detector energy level due to its coupling to the leads, and $\varepsilon_m$ is its level energy relative to the lead chemical potential, so $\varepsilon_m-\lambda/2$ and $\varepsilon_m+\lambda/2$ are the detector level energies for $N_d=0$ and $N_d=1$, respectively.

The strength of AOC backaction is quantified by the AOC exponent $\alpha=(\delta/\pi)^2$: larger $\alpha$ corresponds to smaller overlap between the detector many-body states associated with $N_d=0$ and $N_d=1$.  When an electron tunnels onto or off the dot, the detector scattering potential changes suddenly, projecting the detector Fermi sea from the many-body state appropriate to the initial dot charge onto states of a new detector Hamiltonian.  This many-body response provides a mechanism for inelastic processes that is directly observable in the data, and is encoded in the overlap amplitude $A^{\pm}(t)$ between the initial and final many-body states of the detector, associated with the addition ($+$) or removal ($-$) of an electron from the dot.  The Fourier transforms, $A^{\pm}(E)$, describe the distribution of energy exchanged with the detector during tunnelling.  When $\alpha=0$, $A^{\pm}(E)$ are delta functions, corresponding to purely elastic resonant tunnelling.  Finite $\alpha$ converts these delta functions into power-law spectra governed by the same exponent, shifting spectral weight to finite energy and thereby suppressing elastic tunnelling while enabling inelastic processes.

When a voltage bias, $-V_d\equiv (\mu_L-\mu_R)/e$, is applied across the dot (Fig.~\ref{fig:1}a), the AOC-modified spectral functions $A^{\pm}(E)$ determine the energy-integrated tunnel rates into and out of the dot, and therefore its average occupation.  Following the convention of Ref.~\cite{Nozieres1969}, $A^+(E)$ describes energy $E$ emitted into the detector during tunnel-in, while $A^-(E)$ describes energy $E$ absorbed from the detector during tunnel-out.  At low temperature, with the detector held near equilibrium, energy exchange is directional: $A^+(E)=0$ for $E<0$ and $A^-(E)=0$ for $E>0$.  Without AOC, the tunnel rates are  set only by the bare couplings $\Gamma_{L,R}$, independent of \ed.  With AOC, spectral weight is redistributed to finite energy, producing an asymmetry between tunnel-in and tunnel-out rates that depends on the position of \ed within the bias window and affects the average dot occupation.  As \ed is scanned across this window, the resulting occupation profile provides a direct experimental measure of the AOC exponent $\alpha$.

Figure 1b shows the mesoscopic implementation: a pair of capacitively-coupled quantum dots defined in a GaAs two-dimensional electron gas. Each dot is coupled to independent source and drain leads and tuned into the few-electron regime using electrostatic gates.  The experiment is carried out in a large in-plane magnetic field such that the Zeeman splitting, $g\mu_B B$, exceeds the applied bias and only one spin direction participates in transport.  The dot (orange) is tuned to weak, symmetric coupling, $\Gamma_{L}=\Gamma_{R}\equiv\Gamma_d \ll k_BT$, while the detector (blue) is tuned to strong coupling, $\Gamma_{m}>k_BT$. Linear combinations of the voltages applied to P and DP define virtual gates that control \ed and \eq independently~\cite{supplement}. The detector serves both to induce AOC modifications in the dot and to measure its charge.  A $5~ \mu$V bias applied across the detector enables charge readout while remaining sufficiently weak that classical backaction is negligible~\cite{supplement}.

The charge stability diagram for the dot-detector system, measured with small biases across \emph{both} dots, is shown in Fig.~\ref{fig:1}c.  Simultaneous measurements of the detector current, $I_{m}$ (main panel), and dot current, $I_{d}$ (inset), show the coupled $N_d$ and $N_m$ charge transitions as functions of the two virtual gates controlling \ed and \eq. In Fig.~\ref{fig:1}c and throughout the figures, we plot against $-\ed$ and $-\eq$, chosen so that increasing values correspond to increasing gate voltage; in the text we refer to the corresponding energies $\ed$ and $\eq$.  The detector exhibits a broad Coulomb resonance versus \eq, whose centre shifts  sharply when an electron is added to the weakly-coupled dot.  Transport through the dot (Fig.~\ref{fig:1}c inset) reflects a sharp Coulomb peak versus \ed, whose centre shifts smoothly when the detector charge changes across its strongly-coupled resonance.
The relevant energy scales are extracted directly from the charge stability diagram: the detector broadening, $\Gamma_{m}$, is determined from the width of the detector Coulomb resonance, while the capacitive shift yields $\lambda$. Together, $\Gamma_m$,  $\lambda$ and \eq determine the scattering phase shift in Eq.~\ref{eq:AOC_delta}, and thus the strength of measurement backaction.

Figure~\ref{fig:tr} shows a direct measurement of AOC-modified tunnelling rates and their effect on the average occupation, $\langle N_d \rangle (\ed)$.  A large bias is applied across the dot to expose a wide energy window over which tunnelling processes can be monitored, while $\Gamma_{d}$ is made small enough that individual tunnelling events are resolved within the  $\approx 1$~kHz bandwidth of the charge-sensing circuit.  The result is the detector current map in Fig.~\ref{fig:tr}a, which should be compared to the near-equilibrium stability diagram in Fig.~\ref{fig:1}c where the dot bias is small and $\Gamma_{d}$ is much larger.  Tunnel-in and tunnel-out events are visible as $\ed$ is swept across the bias window, both in the 2D map (Fig.~\ref{fig:tr}a) and in the slow line trace shown in Fig.~\ref{fig:tr}b. For each pair of $\{\ed,\eq\}$, the average occupation is extracted from histograms of \ics, which discriminate the $N_d=0$ and $N_d=1$ charge states (Fig.~\ref{fig:tr}c).
The resulting $\langle N_d \rangle (\ed,\eq)$ are presented as markers in Fig.~\ref{fig:tr}d.

Varying the detector level, \eq, for fixed $\Gamma_m$ and $\lambda$ directly tunes the AOC strength (Eq.~\ref{eq:AOC_delta}), with $\alpha$ maximum near $\eq=0$.  We first consider the outermost traces in Fig.~\ref{fig:tr}d, corresponding to large $|\eq|$.  Here the detector is far off resonance, so the scattering phase shift barely changes when an electron enters the dot, AOC is vanishingly weak and tunnel rates are energy independent; Eq.~\ref{eq:AOC_delta} gives $\alpha\approx 0.002$.  As expected, $\langle N_d \rangle(\ed)$ is flat as \ed traverses the bias window, with a value near 0.5 reflecting the near-equality of $\Gamma_L$ and $\Gamma_R$.  As $|\eq|$ decreases, the plateau in $\langle N_d \rangle(\ed)$ is gradually lost.  Near $\eq=0$, where the scattering phase shift changes maximally with dot occupation and (via Eq.~\ref{eq:AOC_delta}) $\alpha\approx 0.55$, $\langle N_d \rangle(\ed)$ instead evolves almost linearly from 0 to 1 across the bias window.  This evolution is the expected signature of AOC: spectral weight redistributed to finite energy makes the available tunnel-in and tunnel-out processes depend on the position of \ed within the bias window, lowering the average occupation near the top of the window and raising it near the bottom.  The inset of Fig.~\ref{fig:tr}d demonstrates the corresponding evolution of $\alpha$.  The dashed line shows the analytic prediction of Eq.~\ref{eq:AOC_delta}; markers show susceptibility-based estimates obtained from both the measured and calculated occupation traces~\cite{supplement}.  The calculated estimates include the finite-temperature and finite-bias conditions of the experiment, which account for their deviation from the analytic limit.

\begin{figure*}[t]
    \includegraphics[width=\textwidth]{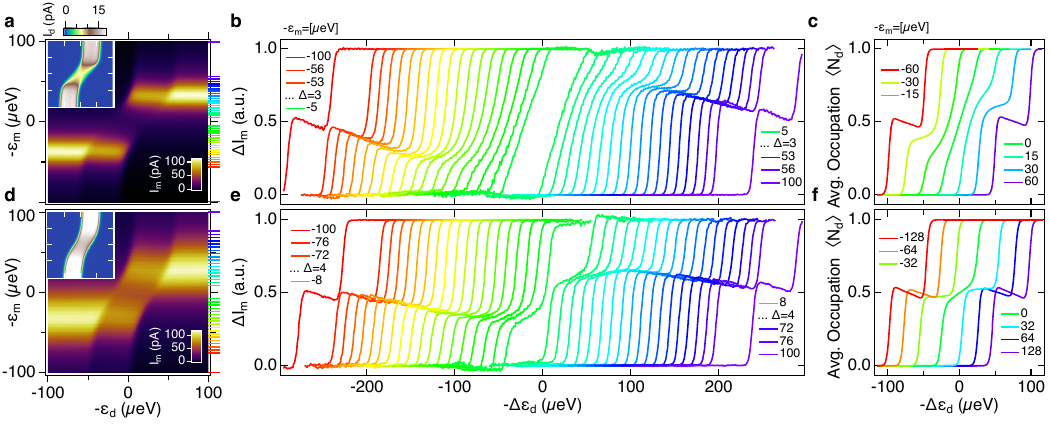}
    \caption{\textbf{Average-current proxy for AOC-modified occupation.} \textbf{a}, Simultaneous detector- and dot-current measurements, showing $I_{m}$ and $I_{d}$ (inset) for $\lambda=70 \ \mu e$V, $\Gamma_m=15 \ \mu e$V, and $V_d=50 \ \mu $V.  The dot coupling $\Gamma_d$ is tuned large enough that individual tunnelling events are no longer resolved.  \textbf{b}, Detector current $I_{m}$ normalized to the values corresponding to $N_d=0$ and $N_d=1$, as a proxy for $\langle N_d\rangle$.  Traces are shown for several detector detunings, \eq, indicated by colour in panel \textbf{a}. \textbf{c}, Calculated $\langle N_d \rangle (\ed)$ for the same parameters, including AOC and an FES. \textbf{d-f}, Equivalent measurement and calculation for parameters corresponding to weaker AOC, $\lambda=60 \ \mu e$V, $\Gamma_m=32 \ \mu e$V.}
    \label{fig:scd}
\end{figure*}

For comparison, Fig.~\ref{fig:tr}e shows theoretical calculations following the approach of Ref.~\cite{Sankar2025Direct}, using parameters determined independently from the charge stability diagram. The theory reproduces the observed evolution of $\langle N_d \rangle (\varepsilon_d)$, from a flat plateau in the weak-AOC limit to an approximately diagonal trace when the detector is near resonance, although the same lineshapes occur at somewhat different values of $\eq$ in experiment and theory.  A second feature, evident in both theory and experiment, is that the occupation near the centre of the bias window shifts away from $ \langle N_d \rangle =0.5$ as $\varepsilon_m$ approaches zero.  This shift reflects an imbalance between tunnel-in and tunnel-out processes, or equivalently between $A^+(E)$ and $A^-(E)$, even when $\ed=0$. The imbalance arises because the detector Hamiltonian itself depends on the dot charge. For tunnelling into the dot, $A^+(E)$ describes a quench from the detector configuration appropriate to $N_d=0$, with detector level $\varepsilon_m-\lambda/2$, to that appropriate to $N_d=1$, with detector level $\varepsilon_m+\lambda/2$. For tunnelling out, $A^-(E)$ describes the reverse quench. Although the AOC exponent is symmetric under the reversal of the quench, the energy-resolved spectra need not be: away from particle-hole symmetry in the detector, $\varepsilon_m=0$, the initial state and final-state manifold correspond to different detector Hamiltonians.  The resulting asymmetry leads to unequal tunnel-in and tunnel-out rates even for equal bare tunnel barriers.

The time-domain signals, such as the trace in Figure~\ref{fig:tr}b, can also be used to extract tunnel-in and tunnel-out rates separately. The average dwell times at $N_d=0$ or $1$ give the corresponding effective tunnel rates, $\Gamma_{in}=\hbar/\tau_{0}$ and $\Gamma_{out}=\hbar/\tau_{1}$. Figure~\ref{fig:tr}f shows $\tau_0(\ed)$ and $\tau_1(\ed)$, for  detector settings corresponding to strong and weak AOC backaction. In the weak-AOC regime, the dwell times are nearly independent of \ed within the bias window, as expected for energy-independent tunnelling.  For strong AOC, the dwell times vary strongly with \ed, confirming directly that AOC backaction appears as energy-dependent tunnel rates into and out of the dot.

The solid lines in Fig.~\ref{fig:tr}d show that $\langle N_d\rangle$ can also be obtained from the time-averaged detector current, $\langle \ics\rangle$, normalized to the values of \ics corresponding to $N_d=0$ and $N_d=1$.  In the time-resolved regime, this average-current proxy agrees closely with the occupation extracted independently from detector-current histograms. The mapping survives nonlinearity in the detector because its response time, $\sim\hbar/\Gamma_m$, is far shorter than the dot tunnelling time: the dot is essentially always in either the $N_d=0$ or $N_d=1$ state, never in between.  This validation makes it possible to explore AOC-modified tunnelling at much larger $\Gamma_d$, in the regime where a typical transport measurement would be carried out.

Figure~\ref{fig:scd} shows the result of such a measurement.  Discrete dot tunnel events are no longer visible in the raw \ics maps (Figs.~\ref{fig:scd}a,d), but each line trace $\ics(\ed)$ corresponding to a fixed value of \eq can be rescaled with respect to the values at large negative and positive \ed.  The resulting normalized signal provides a proxy for $\langle N_d\rangle (\ed)$.  Panels a-c represent a setting of $\lambda/\Gamma_m$ similar to that in Fig.~\ref{fig:tr}, and the primary features of Fig.~\ref{fig:tr}d are indeed reproduced: far from detector resonance the normalized current has a broad plateau, while near $\eq=0$ the plateau is replaced by an approximately-diagonal trace across the bias window.

Beyond what was seen in the time-resolved data, the outer traces show a weak negative slope within the bias window, even when the detector is far from resonance and AOC is negligible.  This behaviour is consistent with 
the closely related Fermi edge singularity (FES), arising from the interaction between an electron tunnelling into the dot and the hole left behind in the lead~\cite{mahan1967excitons,geim1994fermi,abanin2004tunable,frahm2006fermi,krahenmann2017fermi}. The FES enhances the tunnelling density of states near the Fermi energy of the lead, modifying the relative tunnel-in and tunnel-out rates across the bias window.  When the dot level lies near the source Fermi energy, tunnel-in rates are enhanced but tunnel-out rates into the drain are largely unaffected; near the drain Fermi energy the imbalance is reversed.  The result is a negative slope in the average occupation, even in the weak-AOC traces.  This FES contribution accounts for the edge-trace negative slopes, while the much stronger positive slopes near $\eq=0$ reflect AOC backaction from the detector.

Calculations that include both AOC and a Fermi edge singularity are shown in Fig.~\ref{fig:scd}c.  For these calculations, the strength of the FES was adjusted to match the data in the weak-AOC cases at large $|\eq|$, while the remaining parameters ($\lambda,\Gamma_m,\eq,\ed$) were determined independently from the charge stability diagram in Fig.~\ref{fig:scd}a.  With this single additional contribution, the calculation matches the data across the full parameter space.  We speculate that the FES is not apparent in the time-resolved data of Fig.~\ref{fig:tr} because dot-lead coupling is much weaker in that tuning, weakening the FES response.

Figures~\ref{fig:scd}d-f show an analogous measurement for larger detector broadening, $\Gamma_m$, which reduces the change in scattering phase shift and therefore weakens the AOC signature.  The outer traces, where \eq is larger and AOC is negligible, are similar to those in Figs.~\ref{fig:scd}b, as expected.  The weak negative slope from the FES is set primarily by the dot-lead interaction and is not affected by $\Gamma_m$.  Near $\eq=0$, however, the normalized detector current retains a partial plateau rather than becoming fully diagonal, reflecting the weaker AOC-induced energy dependence of the tunnel rates.  Quantitatively, larger $\Gamma_m$ and smaller $\lambda$ in panels d-f imply a maximum $\alpha=0.23$, compared to $\alpha=0.55$ in panels a-c.

\begin{figure}[t]
    \includegraphics[width=\columnwidth]{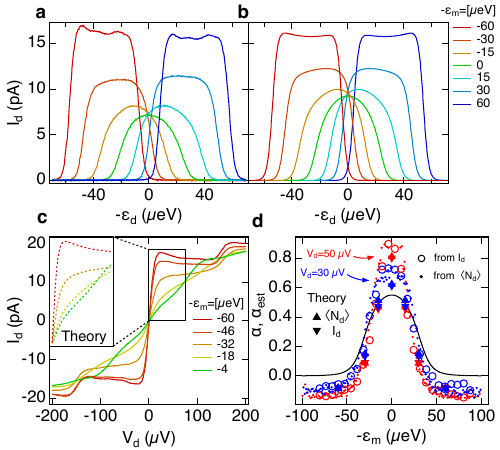}
    \caption{\textbf{Signatures of AOC in current through the dot.} \textbf{a}, Line cuts from Fig.~\ref{fig:scd}a (inset) highlight energy-dependent transmission in the AOC regime. \textbf{b}, Calculation of $ I_d (\ed)$ for the same \eq shown in \textbf{a}, following Ref.~\cite{Sankar2025Direct}, in arbitrary units but consistent across all \eq. \textbf{c}, $I_d(V_d)$ characteristics for various $\eq$, with $\ed$ fixed in the middle of the bias window. Inset: Calculation of $ I_d (V_d)$ (a.u.) for the same parameters shown in \textbf{b} and the bias range indicated in the main panel~\cite{Sankar2025Direct}. \textbf{d}, AOC exponent extracted from $dI_d/dV_d$ (markers) and $d\langle N_d\rangle/d\ed$ (dots), measured at two values of $V_d$, compared with Eq.~\ref{eq:AOC_delta} (solid line). All measurements and calculations correspond to the parameters in Fig.~\ref{fig:scd}a.}
    \label{fig:4}
\end{figure}

Finally, in Fig.~\ref{fig:4} we show how the same AOC-modified tunnel rates appear directly in the source-drain current through the dot,  $I_{d}$, recorded simultaneously with the detector current in Fig.~\ref{fig:scd}a.   In the non-interacting picture,  $I_{d}$ would be independent of \ed while the dot level  is within the bias window,  assuming that excited spin or orbital states remain outside the window, as they do for the measurements presented here~\cite{supplement}.  As shown in Fig.~\ref{fig:4}a, the $I_{d}(\ed)$ traces corresponding to weak AOC ($|\eq|=60~\mu$eV) are indeed nearly flat for \ed within the $50~\mu$eV bias window, with small upturns at the edges likely due to the FES.  Near $\eq=0$, however, the traces become strongly rounded, reflecting the AOC-induced energy dependence of the tunnel rates.
The corresponding calculation, using parameters determined from the charge stability diagram, is shown in Fig.~\ref{fig:4}b. 

Figure~\ref{fig:4}c provides a complementary view of the same physics through tunnelling spectroscopy. Here $I_d$ is measured as a function of source-drain bias $V_d$, with the dot level fixed in the middle of the bias window, directly probing the energy-dependent transmission for several detector detunings \eq. 
In the weak-AOC limit, with the detector far from resonance, the current shows a sharp thermally-broadened onset followed by a plateau, as expected for elastic tunnelling through a single dot level with $A^{\pm}(E)\rightarrow \delta(E)$. 
As $\abs{\eq}$ is reduced and $\alpha$ increases, this plateau progressively disappears: spectral weight is redistributed to finite energy in $A^{\pm}(E)$, allowing inelastic transport channels to contribute as the bias window opens.
Near $\eq =0$, where the scattering phase shift is maximized and $\alpha\approx0.55$, $I_d$ rises continuously across the full accessible bias range, signalling a transport regime in which inelastic, many-body-mediated processes dominate. The corresponding calculation in Fig.~\ref{fig:4}c (inset), using parameters determined independently from the charge stability diagram, reproduces this evolution.

Two additional spectroscopic features are visible in the data. At large bias, $V_d \approx \pm150 \ \mu $V, the weak-AOC traces display a step at the onset of transport through the Zeeman-split excited spin state at $\bpar=3 \ $T. This feature progressively washes out as AOC strengthens. In addition, the far-detuned trace ($\varepsilon_m=60 \ \mu e$V) exhibits a slight enhancement of $I_d$ near threshold, consistent with an FES-enhanced tunnelling density of states at the dot-lead Fermi level. 

Figure~\ref{fig:4}d summarizes the effective AOC exponent extracted as a function of detector detuning, \eq, from both the charge-sensing measurements of Fig.~\ref{fig:scd}b and the transport spectroscopy shown in Fig.~\ref{fig:4}c. For the transport data, $\alpha$ is obtained by fitting the measured $I_d(V_d)$ characteristics to the theoretical model of Ref.~\cite{Sankar2025Direct}. The solid line shows the prediction of Eq.~\ref{eq:AOC_delta} using independently determined values of $\lambda$ and $\Gamma_m$. The exponents extracted from the two experimental signatures ~\cite{supplement} agree closely with one another and follow the expected dependence on detector detuning, while the apparent negative values obtained when the detector is far detuned reflect the competing FES contribution.

We have experimentally isolated Anderson orthogonality as a controllable mechanism of measurement backaction in a capacitively coupled quantum-dot system.  By tuning the detector energy level, we control the scattering phase-shift change induced by dot occupation and thereby tune the strength of the many-body backaction.
The resulting modification of tunnelling dynamics is observed through three complementary signatures. Time-resolved charge sensing reveals the energy dependence of individual tunnel-in and tunnel-out rates; average-occupation measurements capture the integrated effect of this asymmetry across the nonequilibrium bias window; and transport spectroscopy demonstrates the corresponding evolution of dot transmission, from flat profiles consistent with resonant transport when AOC is weak, to line shapes indicating the dominant role of inelastic processes when AOC is maximal. Across these measurements, the experimental evolution agrees with theoretical predictions over a broad range of parameters.  These results establish Anderson orthogonality as a measurable and controllable form of quantum measurement backaction in mesoscopic solid-state devices, and provide a framework for studying how many-body detector environments govern quantum system dynamics beyond the conventional paradigm of dephasing and classical noise.  More broadly, the ability to induce strong many-body backaction offers a route to studying monitored quantum systems in regimes where measurement itself can drive collective changes of state, including measurement-induced phase transitions~\cite{Goldstein_2010,Ma2023Identifying}.

\section{Acknowledgements}
Experiments were undertaken with support from the Natural Sciences and Engineering Research Council of Canada; the Canada Foundation for Innovation; the Canadian Institute for Advanced Research; the Max Planck-UBC-UTokyo Centre for Quantum Materials and the Canada First Research Excellence Fund, Quantum Materials and Future Technologies Program; and the European Research Council (ERC) under the European Union’s Horizon 2020 research and innovation program, Grant Agreement No. 951541. YM acknowledges support from the ISF Breakthrough Grant no. 737/24.  Work in the Manfra group at Purdue University was supported by the US DOE Office of Basic Energy Sciences under Award DE-SC0020138.

\section{Author Contributions}

WG led the measurements, assisted by EC, VM, JD and SL, and supervised by JF.  SS proposed the experiment and developed the theory, under guidance from YM and ES.  SF and GG grew the two dimensional electron gas under supervision of MM.  WG and JF wrote the text with assistance from all authors.  

\section*{Competing interests}
The authors declare no competing interests.

\section*{Additional Information}
Correspondence and requests for materials should be addressed to Joshua Folk.

\section*{Data Availability}
Source data are available for this paper upon request.

\bibliographystyle{apsrev4-1}
\bibliography{main}
\newpage
\clearpage

\onecolumngrid 
\raggedbottom
\setcounter{figure}{0}
\renewcommand{\thefigure}{S\arabic{figure}}
\setcounter{equation}{0}
\renewcommand{\theequation}{S\arabic{equation}}
\setcounter{secnumdepth}{3}
\setcounter{section}{0}
\renewcommand{\thesection}{S\arabic{section}}
\renewcommand{\thesubsection}{S\arabic{section}.\arabic{subsection}}
\renewcommand{\thesubsubsection}{S\arabic{section}.\arabic{subsection}.\arabic{subsubsection}}

\makeatletter
\renewcommand\section{\@startsection{section}{1}{\z@}%
  {-3.5ex \@plus -1ex \@minus -.2ex}%
  {2.3ex \@plus.2ex}%
  {\raggedright\normalfont\Large\bfseries}}
\renewcommand\subsection{\@startsection{subsection}{2}{\z@}%
  {-3.25ex\@plus -1ex \@minus -.2ex}%
  {1.5ex \@plus .2ex}%
  {\raggedright\normalfont\large\bfseries}}
\renewcommand\subsubsection{\@startsection{subsubsection}{3}{\z@}%
  {-3.25ex\@plus -1ex \@minus -.2ex}%
  {1.5ex \@plus .2ex}%
  {\raggedright\normalfont\normalsize\bfseries}}
\makeatother

\begin{center}
\Large\textbf{Methods:}\\Anderson Orthogonality as Measurement Backaction in Coupled Quantum Dots
\end{center}

\section{Theoretical Model} \label{sec:model}
The theoretical calculations presented in the paper are based on Ref.~\onlinecite{Sankar2025Direct}. Here, for the sake of completeness, we discuss the considered theoretical model. 
The combined system plus detector is described by the Hamiltonian,
\begin{equation}
\label{eq:H_tot}
   H=H_{\rm{{sys}}}+H_{\rm{{det}}}+H_\lambda.
\end{equation}
$H_{\rm{{sys}}}$ describes the spinless system quantum dot (QD) tunnel coupled to voltage biased leads (at potentials $\mu_{L,R}$),
\begin{equation}
\label{eq:H_sys}
    H_{\rm{{sys}}} = H_{\rm{{dot}}}+H_{\rm{{leads}}}+H_{\rm{{u}}}+H_{\rm{{tun}}},
\end{equation}
with,
\begin{equation}
    H_{\rm{{dot}}}= \epsilon_{d} d^\dagger d,\quad  H_{\rm{{leads}}}= \sum_{i=L,R}\sum_k  \epsilon_{k} c_{ik}^\dagger c_{ik}, \quad H_{\rm{{tun}}}=  \sum_{i=L,R}(\gamma_i c_{i1}^\dagger d+ H.c.), \quad H_{u}=  \sum_{i=L,R}u_i(c_{i1}^\dagger c_{i1})(d^\dagger d). 
\end{equation}
Here $d^\dagger$ creates an electron in the QD,  $c_{ik}^\dagger$ creates an electron with momentum $k$ in  lead $i$, and $c_{i1}=\sum_k c_{ik}$. The tunneling line width is $\Gamma_{i}=2\pi \nu_0 \gamma_i^2$, where $\nu_0$ is the density of states in the leads.
 We consider two charge states of the QD denoted $n=0,1$. Since we only consider the limit of weak tunneling in the QD, $\Gamma_{\rm{L},\rm{R}} \ll T$,  hybridization effects due to tunneling are neglected. 
$H_u$ describes the capacitive interaction between the dot and leads. This interaction serves as an $n-$ dependent scattering potential for the leads and results in the Fermi edge singularity (FES) discussed in the paper. For later purpose, we define $n-$ dependent  Hamiltonian of the leads as, $H^{\rm{{lead,\,i}}}_n= H_{\rm{{lead,\,i}}}+H_{u,\,i} $.

The detector  consists of another quantum dot (QDD). Its Hamiltonian including its coupling with the system's QD depends on $n$, and is described by $H^{\rm{{det}}}_n = H_{\rm{{det}}}+H_\lambda$ with
\begin{align}
\label{eq:H_det}
H^{\rm{{det}}}_n&=\epsilon_{{\rm{m}}} f^\dagger f +\sum_{\mu=L,R} \sum_k(\epsilon_k \varphi_{k\mu}^\dagger \varphi_{k\mu} + v_\mu \varphi_{k\mu}^\dagger f +h.c. )  +\lambda\left(f^\dagger f-\frac{1}{2}\right)\left(n-\frac{1}{2}\right),
\end{align}
where $f^\dagger$ creates an electron in the QDD, and $\varphi_{k\mu}^\dagger$ creates an electron with momentum $k$ in detector lead $\mu$.
The total tunneling width of the QDD is $\Gamma_{\rm{m}}=2\pi \nu_m (v_L^2+v_R^2)$, where $\nu_m$ is the density of states in the leads. The tunneling in the QDD is assumed to be strong, $\Gamma_{\rm{m}}\gg T$.

 The tunnel rates of the system QD from/to the lead $i=L/R$ are given by,
 \begin{eqnarray}
    \Gamma_{i,\,in}(\epsilon_d)&=&\gamma_i^2\int_{-\infty}^{\infty}dE\, \hat{B}_i^+(\mu_i-\epsilon_d-E)\hat{A}^+(E),\label{eq:tun_in_fes_det}\\
     \Gamma_{i,\,out}(\epsilon_d)&=& \gamma_i^2\int_{-\infty}^{\infty}dE\, \hat{B}_i^-(\mu_i-\epsilon_d-E)\hat{A}^-(E),\label{eq:tun_out_fes_det}
\end{eqnarray}
where the functions, $\hat{B}_i^{\pm}(E)$ and  $\hat{A}^{\pm}(E)$ are respectively the Fourier transform of the correlators $B_i^{\pm}(t)$ and $A^\pm(t)$, defined as,
\begin{equation}
    B_i^+(t)={\rm{Tr}}\left[\rho^{\rm{{lead,\,i}}}_0 c_{i1}^\dagger(t) e^{itH^{\rm{{lead,\,i}}}_0}e^{-itH^{\rm{{lead,\,i}}}_1} c_{i1}(0)\right], \quad B_i^-(t)={\rm{Tr}}[\rho^{\rm{{lead,\,i}}}_1 c_{i1}(0)e^{itH^{\rm{{lead,\,i}}}_0}e^{-itH^{\rm{{lead,\,i}}}_1}c_{i1}^\dagger(t)],
\end{equation}
\begin{equation}
    A^+(t)={\rm{Tr}}\left[\rho^{\rm{{det}}}_0 e^{itH^{\rm{{det}}}_0}e^{-itH^{\rm{{det}}}_1}\right],\quad  A^-(t)={\rm{Tr}}\left[\rho^{\rm{{det}}}_1 e^{itH^{\rm{{det}}}_0}e^{-itH^{\rm{{det}}}_1}\right].
\end{equation}
Here $\rho^{\rm{{lead,\,i}}}_n \propto e^{- H^{\rm{{lead,\,i}}}_n/T} $ and $\rho^{\rm{{det}}}_n \propto e^{- H^{\rm{{det}}}_n/T} $  respectively denote the system leads and detector density matrix for QD charge $n$. 

The interaction strengths $u_i$ control the FES exponent $\alpha'$ that determines the low-energy behavior of $\hat{B}_i^\pm(E)$. Assuming identical interaction strengths, $u_L=u_R$, we drop the subscript $i$. The low energy behavior of $\hat{B}^\pm(E)$ is given by,
\begin{equation}
\label{eq:bpm_analytical}
    \hat{B}^\pm(E)=B\nu_0 \Gamma(\alpha')  T^{-\alpha'}{\rm{Re}}\left[\frac{e^{\pm i(\pi/2)(1-\alpha')}\Gamma\left(\frac{iE}{2\pi T}+\frac{1-\alpha'}{2}\right)}{\Gamma\left(1+\frac{iE}{2\pi T}-\frac{1-\alpha'}{2}\right)}\right] \quad \rm{for}\quad |E|\ll D,
\end{equation}
where
$\Gamma(\cdot)$ denotes the gamma function, $B$ is a non-universal constant and $D$ is some high energy cutoff associated with the system leads. The interaction strength $\lambda$ controls the AOC exponent $\alpha$ that determines the low-energy behavior of $\hat{A}^\pm(E)$,
\begin{equation}
\label{eq:apm_analytical}
    \hat{A}^\pm(E)=\Gamma(1-\alpha)\frac{A}{D}\left(\frac{T}{D}\right)^{\alpha-1}{\rm{Re}}\left[\frac{e^{\pm i\pi\alpha/2}\Gamma\left(\frac{iE}{2\pi T}+\frac{\alpha}{2}\right)}{\Gamma\left(1+\frac{iE}{2\pi T}-\frac{\alpha}{2}\right)}\right] \quad \rm{for}\quad |E|\ll D,
\end{equation}
where $A$ is a non-universal constant and  $D$ is again some high energy cutoff associated with the detector bandwidth.

In the paper, for the tuning where time resolved data is obtained, the FES signatures are not observed. This tuning corresponds to a very weak tunnel coupling between the QD and leads and we believe that this consequently reduces the capacitive coupling as well. Thus for this tuning, $u_i \to 0$ in the Hamiltonian and consequently $\alpha' \to 0$. It then follows from Eq.~\eqref{eq:bpm_analytical} that,
\begin{equation}
    \hat{B}^\pm(E) \propto \nu_0 f(\mp E),\quad {\rm{for}}\quad \alpha'\to 0,
\end{equation}
where $f(\cdot)$ denotes the Fermi function. Then the tunnel rates in Eqs.~\eqref{eq:tun_in_fes_det} and \eqref{eq:tun_out_fes_det} become
 \begin{eqnarray}
    \Gamma_{i,\,in}(\epsilon_d)&=&\Gamma_i\int_{-\infty}^{\infty}dE\, f(E-\mu_i+\epsilon_d)\hat{A}^+(E),\label{eq:tun_in_det}\\
     \Gamma_{i,\,out}(\epsilon_d)&=& \Gamma_i\int_{-\infty}^{\infty}dE\, (1-f(E-\mu_i+\epsilon_d))\hat{A}^-(E).\label{eq:tun_out_det}
\end{eqnarray}

\section{Experimental Methods}\label{sec:methods}

\subsection{Device Fabrication}

The device consists of a pair of capacitively coupled GaAs quantum dots defined electrostatically in a GaAs/AlGaAs heterostructure hosting a two-dimensional electron gas (2DEG) located 57\,nm below the surface. The heterostructure had a carrier density and mobility at 300\,mK of $2.42\times10^{11}\,\mathrm{cm}^{-2}$ and $2.56\times10^{6}\,\mathrm{cm}^2/(\mathrm{V\,s})$, respectively, determined in a separate calibration measurement. Mesas were defined using UV laser lithography, followed by electron beam lithography and deposition of NiAuGe ohmic contacts to the 2DEG. Following ohmic fabrication, 10\,nm of HfO$_2$ was deposited by atomic layer deposition to improve electrostatic gate stability and reduce charge noise. Electrostatic gates defining the quantum dot circuit were fabricated using a two-stage electron beam lithography process followed by electron beam evaporation. A fine gate step was used for the inner gate structures, while a coarse gate step defined the larger outer gate structures and bond pads. In the fine step, 2/12\,nm of Pd/Au were deposited. In the coarse step, 10/150\,nm of Ti/Au were deposited. Electrical connections between the sample gates, ohmic contacts, and chip carrier were made using Al bond wires. A scanning electron micrograph of the fabricated device is shown in the schematic of Fig.~\ref{fig:1}b. Metallic top gates (light grey) electrostatically define the coupled quantum dot system within the 2DEG, and tune the tunnel barriers and dot chemical potentials. Squares denote regions containing ohmic contacts to the electron gas. The grounded gate shown in Fig.~\ref{fig:1}b is used to provide additional screening to limit cross-capacitive interaction.

\subsection{Measurement Configuration}

Transport measurements were carried out in a Bluefors XLD dilution refrigerator with a nominal base mixing chamber temperature of 10\,mK and a one-axis superconducting magnet. An in-plane magnetic field of $B=3\,\mathrm{T}$ was applied throughout all measurements.
All measurement and control electronics were located at room temperature and connected to the device through filtered DC lines. Each line was low-pass filtered using inline RC filters mounted at the mixing chamber stage with a cutoff frequency of approximately 16\,kHz, to suppress high-frequency electrical noise.
Gate voltages and DC biases were applied using a custom-built synchronized DAC/ADC measurement platform optimized by our group and the UBC Physics and Astronomy technical staff, based on the open-source OpenDAC architecture (\url{https://opendacs.com/}) developed by Hugh Churchill and Andrea Young. The system was implemented using an Arduino Due interfaced with Analog Devices AD5764 DAC and AD7734 ADC evaluation boards. Modifications were made to the hardware configuration and acquisition firmware, including synchronization of DAC sweeps and continuous acquisition of ADC signals. Gate voltages were updated at 9.709 kHz, with ADC acquisition synchronized to each DAC update. This ensured deterministic sampling of the transport and charge sensing signals ($I_d$, $I_m$) throughout each sweep and eliminated timing uncertainty between gate updates and current acquisition. Some gate electrodes were driven by two independently addressable DAC channels connected through respective voltage dividers, the outputs of which were summed at the device. This provided independent coarse and fine control of the electrostatic potential while maintaining high effective voltage resolution. $I_d$ and $I_m$ were measured simultaneously using Basel SP983c transimpedance amplifiers, configured with a gain of $10^{9}\,\mathrm{V/A}$, and  3\,kHz bandwidth set by a two-stage low-pass filter. For measurements in the average-current regime, $I_d$ and $I_m$ are digitally notch filtered at 60, 180, and 300 Hz (quality factors Q=1, 3, 5, respectively) to remove line-frequency pickup and its harmonics, then resampled from the native 9.709 kHz acquisition rate down to 300 Hz. The time-resolved data are left unfiltered, since notch filtering would distort sharp edges of individual tunnelling events; instead they are resampled by a separate factor chosen so the bimodal ADC distribution cleanly separates the $N_d=0$ and $N_d=1$ states without inflating the count of detected switching events entering the dwell-time statistics. 

\subsection{System Characterization}

 \subsubsection{Tunnel Barrier Symmetrization}

 The system dot and detector were first tuned to the single-electron regime. Barrier gate voltages (LC, RC, DLC, DRC) controlled the tunnel rates to the source and drain leads, while plunger gates (P, DP) controlled the dot level energies. Next, the tunnel barriers of both the dot and the detector were symmetrized. The detector was symmetrized by examining the transport as a function of the detector barrier gates $(DLC,DRC)$. In this space, the Coulomb resonance of the detector forms a linear contour. The point of maximum resonant transport along this contour was selected as the operating point, assuming symmetric tunnel coupling according to the standard resonant transport relation:
\begin{equation}
I_m \propto \frac{\Gamma_{m,L} \Gamma_{m,R}}{\Gamma_{m,L} + \Gamma_{m,R}},
\end{equation}
which is maximized for $\Gamma_{m,L} = \Gamma_{m,R}$. The detector coupling $\Gamma_m$ was independently determined from the full-width at half maximum (FWHM) of the detector Coulomb resonance (see Sec.~S1.3.3). The system dot was symmetrized independently by examining the charge-sensed transition with the detector strongly detuned from resonance, corresponding to a regime of weak AOC strength where the detector acts as an effectively noninvasive charge sensor. With a finite bias applied across the system dot, the occupation in the centre of the bias window is:
\begin{equation}
\langle N_d\rangle (\ed=\frac{\mu_L+\mu_R}{2})
=
\frac{\Gamma_L}{\Gamma_L + \Gamma_R},
\end{equation}
such that symmetric coupling corresponds to $\langle N_d\rangle= 0.5$. The system barrier gates were tuned to bring the midpoint occupation as close as possible to this condition. The system dot tunnel coupling was tuned to remain significantly below temperature, as indicated by the strongly suppressed conductance through the dot relative to the conductance quantum $e^2/h$.

 \subsubsection{Construction of the Virtual Gate Basis}
\begin{figure*}[t]
    \centering
    \includegraphics[width=0.4\columnwidth]{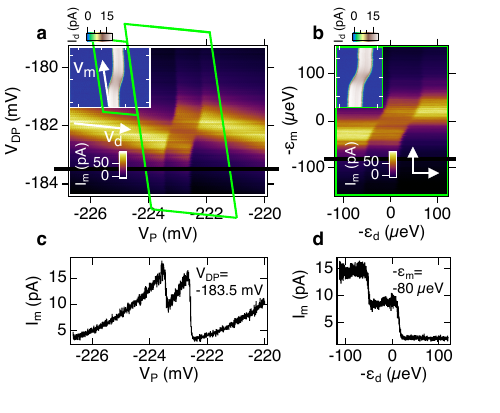}
    \caption{\textbf{Virtual gates.} \textbf{a}, Gate voltage map showing simultaneous measurement of $I_m$ (main panel) and $I_d$ (inset), while sweeping in the real gate space. \textbf{b}, Virtual gate map, in which the Coulomb resonance of the system dot and detector are effectively orthogonal. The extent of the virtual sweep in the real gate space is boxed in green in \textbf{a}. \textbf{c}, Line cut through \textbf{a}, highlighting the extent to which sweeping the system dot plunger cross-capacitively effects the detector dot level. \textbf{d}, Line cut of \textbf{b}, highlighting an orthogonal virtual basis, where sweeping the virtual gate has little effect on the detector current.}
    \label{fig:virtual}
\end{figure*}
 Throughout this experiment we use virtual gates made up of a linear combination of dot and detector plungers in order to shift the energy of one of them without affecting the other, constructing an approximately orthogonal coordinate system of the system and detector dot energies $(\varepsilon_d,\varepsilon_m)$ while compensating for cross-capacitive coupling in the system. To first order, sweeping either of the physical plunger gates $P$ or $DP$ produces two simultaneous effects: a direct lever-arm coupling to the corresponding dot energy, and a cross-capacitive contribution to the other dot. The mapping between gate voltage space and energy space may therefore be written as a linear transformation:
\begin{equation}
\begin{pmatrix}
\delta V_P \\
\delta V_{DP}
\end{pmatrix}
=
\mathbf{M}
\begin{pmatrix}
\delta \varepsilon_d \\
\delta \varepsilon_m
\end{pmatrix},
\label{eq:virtualmap}
\end{equation}
where $\mathbf{M}$ is a matrix contains both the direct lever arms and cross-capacitive couplings.
The virtual basis vectors defining $(\varepsilon_d,\varepsilon_m)$ were extracted geometrically from charge stability diagrams in $(P,DP)$ space. Far from the interdot transition, the Coulomb peak corresponding to one dot traces a contour of that dot's own constant energy, providing the direction that serves as the basis vector for the other dot's virtual energy axis. Consequently, the basis vector corresponding to $\varepsilon_d$ was defined parallel to the detector Coulomb peak, while the basis vector corresponding to $\varepsilon_m$ was defined parallel to the system dot transition. This procedure is illustrated schematically in Fig.~\ref{fig:virtual}. Rather than immediately converting into physical energy units, the virtual basis was first constructed to preserve the metric structure of voltage space. Each basis vector was therefore Euclidean-normalized in the original gate-voltage coordinates such that voltage steps remained uniform in the transformed basis. Explicitly,
\begin{equation}
\mathbf{M}_{\mathrm{affine}}
=
\begin{pmatrix}
v_{d,P} & v_{m,P} \\
v_{d,DP} & v_{m,DP}
\end{pmatrix},
\end{equation}
where
\begin{equation}
v_d
=
\frac{
\begin{pmatrix}
\Delta V_{P,d} \\
\Delta V_{DP,d}
\end{pmatrix}
}{
\sqrt{
(\Delta V_{P,d})^2 + (\Delta V_{DP,d})^2
}},
\end{equation}
and similarly for $v_m$. The affine transformation therefore defines a virtual coordinate system in units of gate voltage rather than energy. If the initial affine basis selection produced small residual nonorthogonalities, the transformation was iteratively refined through an additional fine correction matrix,
\begin{equation}
\mathbf{M}_{\mathrm{affine}}
=
\mathbf{M}_{\mathrm{affine}}^{(\mathrm{coarse})}
\,
\mathbf{M}_{\mathrm{affine}}^{(\mathrm{fine})},
\end{equation}
which was computed by repeating the procedure above within the coarse virtual space. The final composition of $\mathbf{M}$ as defined in Eq.~\ref{eq:virtualmap} includes an appropriate lever arm scaling into energies (from $m$V to $\mu e$V). Conversion into energy units was performed in an additional step through a scaling matrix,
\begin{equation}
\mathbf{M}_{\mathrm{scale}}
=C
\begin{pmatrix}
1 & 0 \\
0 & \lambda_m/\lambda_d
\end{pmatrix},
\end{equation}
where $\lambda_d$ and $\lambda_m$ are the affine space distances of the capacitive shift for the dot and detector Coulomb peaks respectively, and $C$ is an overall scalar calibration factor to fix the \ed basis vector in units of energy. In particular, the applied source-drain bias of $50\,\mu$V defines the extent of the bias window feature, providing an absolute calibration of the $\varepsilon_d$ axis by scaling the window to $50 \ \mu e$V. Then, the rescaling $\lambda_d/ \lambda_m$ calibrates the \eq axis, by the symmetry of the capacitive interaction, since the capacitive shift produced on one dot is equal and opposite to the shift on the other. This matrix also contains the effective lever arm information for the virtual gates. The full transformation is therefore:
\begin{equation}
\mathbf{M}
=
\mathbf{M}_{\mathrm{affine}}
\,
\mathbf{M}_{\mathrm{scale}},
\end{equation}
such that we recover Eq.~\ref{eq:virtualmap}. 

 \subsubsection{Extraction of the Relevant Energy Scales from Charge Stability Diagrams}
 Once the charge stability diagram is expressed in units of energy, the relevant scales for the experiment can be read off directly from the 2D virtual map of $I_m(\ed,\eq)$, such as the one shown in Fig.~\ref{fig:scd}a. As discussed above, the capacitive coupling $\lambda$ is determined from the shift in the Coulomb peak of either system dot or detector. Because the detector is tuned to strong coupling ($\Gamma_m \ \gg k_BT$), its resonance is Lorentzian rather than thermally broadened, and $\Gamma_m$ is determined from the full width at half maximum (FWHM) of the detector peak away from the interdot transition, explicitly as $\Gamma_m=FWHM/2$. 
 An estimate for electron temperature is determined via Coulomb blockade thermometry of the system dot, in particular from fitting the thermally broadened edges of the bias window for the transmission current, giving an electron base temperature $T\approx23 $ mK. We fit to:
\begin{equation}
I_d(\ed)=I_d^{max}[f(\ed-\mu_R)-f(\ed-\mu_L)]+I_d^0,
\end{equation}
where 
\begin{equation}
f(x)=\frac{1}{e^{x/k_BT}+1},
\end{equation}
and $\mu_L-\mu_R=50 \ \mu e$V, the canonical lineshape for sequential tunnelling through a single, well-resolved level between two Fermi edges separated by the calibrated bias window. The in-plane field $\bpar=3 \ $T applied to enforce spinless transport leads to Zeeman-splitting of the excited spin state well outside the bias window. That this condition holds is confirmed directly by Coulomb blockade spectroscopy (Fig.~\ref{fig:4}c, Fig.~\ref{fig:diamonds}b,c), which resolves the excited state only at much larger bias. 

\begin{table}[htbp]
    \centering
    \caption{Summary of device parameters used across measurements.}
    \label{tab:figure_parameters}
    \begin{tabular}{@{}lcc@{}}
        \toprule
        \textbf{Figure(s)} & \textbf{$\lambda$ ($\mu e$V)} & \textbf{$\Gamma_m$ ($\mu e$V)} \\
        \midrule
        Fig.~\ref{fig:tr} & 104 & 22 \\
        Fig.~\ref{fig:1}c, Fig.~\ref{fig:scd}a, b, c, Fig.~\ref{fig:4} & 70 & 15 \\
        Fig.~\ref{fig:scd}d, e, f & 60 & 32 \\
        \bottomrule
    \end{tabular}
\end{table}

\newpage
\section{Time-Resolved Analysis (Fig.~2)}

 \subsection{Occupation Processing}\label{subsec:occprocess}
Two acquisition modes are reflected in Fig.~\ref{fig:tr}. The first, shown in Fig.~\ref{fig:tr}a, is a quick scan sweeping $\varepsilon_d$ rapidly over the charge transition (4 s per row), sufficient to resolve individual tunnelling events visually. The second, which forms the basis of the quantitative analysis presented in Figs.~\ref{fig:tr} b–f, is based on longer time-resolved measurements of the detector current, acquired by stepping $\varepsilon_d$ across the dot transition for multiple detector detunings $\varepsilon_m$.
At a given gate setting, the detector current is recorded for $1\,\mathrm{s}$ before advancing to the next gate step, with each \ed sweep consisting of $150$ gate positions over the dot charge transition. Each \ed sweep was repeated 20 times to improve counting statistics.

The resulting dataset then consists of 20 1~s time-domain traces at 150 different gate positions. First, the data is digitally resampled to 300~Hz to suppress high-frequency noise while preserving the bimodal structure of $N_d=0$ and $N_d=1$. All signal samples corresponding to the same gate position were combined into a single ensemble. Histograms constructed from these ensembles represent the statistics of the detector current as a function of $\varepsilon_d$. This representation converts the time-domain telegraph signal into a probability distribution of detector currents from which the average occupation can be extracted. Fig.~\ref{fig:tr}c shows the evolution of these histograms across the charge transition. Far from the transition, nearly all histogram weight resides in a single state corresponding to either $N_d=0$ or $N_d=1$. Within the bias window set by $V_d$, spectral weight is redistributed between the two states as tunnelling becomes energetically allowed. The occupation probability at each \ed was extracted from these histograms via a thresholding procedure. For each full dataset covering a range of \ed (such as that shown in Fig.~\ref{fig:tr}c), a global histogram was constructed and used to define a threshold current, $I_m^{\mathrm{th}}$, chosen to be the minimum between the two peaks of the bimodal distribution. Using $I_m^{\mathrm{th}}$, the occupation at gate position ($\varepsilon_d$, $\varepsilon_m$) was computed by integrating over the histogram:
\begin{equation}
\langle N_d \rangle(\varepsilon_d,\varepsilon_m)
=
\frac{
\displaystyle
\int_{I_m^{\mathrm{th}}}^{\infty}
P(I_m,\varepsilon_d,\varepsilon_m)\, dI_m
}{
\displaystyle
\int_{-\infty}^{\infty}
P(I_m,\varepsilon_d,\varepsilon_m)\, dI_m
},
\end{equation}
where $P(I_m,\varepsilon_d,\varepsilon_m)$ is the detector current probability distribution  at fixed gate position. Equivalently, $\langle N_d\rangle$ is simply the fraction of time the detector resides in the occupied state.

This process was repeated $\approx$10 times at a given \eq to gather more statistics.  In order to combine these 10 repeats, each taking approximately an hour, they were realigned to compensate for slow offset-charge drift during the experiment. This realignment was done using a one-dimensional cross-correlation. A representative trace near the median displacement served as the reference, and the remaining traces were shifted by an integer number of gate points to maximize the cross-correlation. After alignment, the occupation traces for a given detector setting were averaged together to produce the final occupation curve $\langle N_d \rangle(\varepsilon_d)$. For comparison, the detector current was also averaged directly, with no realignment. As shown in Fig.~\ref{fig:tr}d, this blind average of the detector current reproduces the occupation obtained from the histogram analysis, confirming that it provides an accurate proxy for $\langle N_d\rangle$. 

 \subsection{Extraction of Dwell Time Statistics}

 Average occupation discards information about the underlying switching dynamics: the same telegraph noise data used in Sec.~S2.1 give direct access to the mean dwell time in each charge state as a function of \ed. Fig.~\ref{fig:tr}f shows the dwell time extracted for detector settings corresponding to strong ($-\eq=-6 \ \mu e$V) and weak ($-\eq=74 \ \mu e$V) AOC. For a given \eq, the full set of resampled detector-current values at each gate step (Sec. S2.1) is thresholded using the same global histogram threshold as the occupation analysis, converting the continuous trace into a binary occupied/empty sequence in time. Each run in one state is a single dwell event, of a known duration. At each gate step, $\tau_0(\ed)$ and $\tau_1(\ed)$ are the mean durations of all $N_d=0$ and $N_d=1$ state events, accumulated across every repeat of the sweep. Within the bias window ($\mu_R < \varepsilon_d < \mu_L$), an empty dot can only be filled from the source and an occupied dot can only empty into the drain, so $\tau_0=\hbar/\Gamma_{in}$ and $\tau_1=\hbar/\Gamma_{out}$.

\newpage
\section{Analysis  for Figs.~3 and 4}

 \subsection{Correlation Averaging}
Each average-current dataset (Figs.~\ref{fig:scd},\ref{fig:4}) consists of many repeated 2D sweeps of the simultaneously measured dot and detector currents ($I_d,I_m$).  Multiple quick datasets are collected, then aligned and averaged, instead of taking one slower dataset, in order to account for slow gate drift and occasional offset-charge jumps. Alignment of data before averaging is carried out using $I_d$, as the transport peak through the weakly-coupled dot is sharper and better-defined than the broad resonance in $I_m$; because the two currents are acquired simultaneously at every gate point, whatever shift aligns $I_d$ also aligns $I_m$.

Alignment is performed only along \ed: a comparison of the many repeated 2D scans indicated no discernable shift in \eq. For each \eq value independently, the \ed position of the dot's Coulomb peak is tracked across repeats by one-dimensional cross-correlation, allowing for the possibility that drift is not a single rigid shift over the full \eq range. To avoid anchoring the alignment to one, possibly atypical, repeat, this is done in two passes: the per-\eq peak trajectory is first computed for every repeat and averaged into a robust reference trajectory, and each repeat is then re-aligned onto this average rather than onto any single raw repeat. The resulting shift is applied identically to the $I_d$ and $I_m$ maps for that repeat, and aligned repeats are averaged together (following notch filtering and resampling as in Sec. S1.2) to produce the high signal-to-noise maps used in Figs.~\ref{fig:scd} and \ref{fig:4}.
 \subsection{Occupation Normalization}
 Each \eq column of the aligned, averaged detector-current map is independently rescaled into a curve that reflects occupation, running from 0 to 1, and labelled $I_m$(a.u.). Two reference rows, taken well outside the bias window and hence deep in the established $N_d=0$ and $N_d=1$ plateaus, fix the two endpoints, with the sign chosen so occupation increases with gate voltage regardless of the polarity of change in $I_m$ (i.e. which side of the detector peak is being used to sense charge). Doing this independently for every \eq compensates for the slow variation in the detector's absolute signal size and backgound across the full \eq range. The result is a $\Delta I_m(\ed,\eq)$ spectrum, as shown in Fig.~\ref{fig:scd}b, used directly for the $\alpha$ extraction of Sec. S4.

\section{Extraction of the AOC Exponent (Figs.~2 and 4)}
Estimates of $\alpha$ extracted from experimental parameters are compared to theoretical calculations in Figs.~\ref{fig:tr}d(inset) and \ref{fig:4}d.  The theoretical $\alpha$ itself, from Eq.~\ref{eq:AOC_delta}, is shown as dashed or solid lines in Figs.~\ref{fig:tr}d(inset) and \ref{fig:4}d respectively.  Numerical estimates of this quantity, $\alpha_{est}$ are shown as markers, extracted independently from the average occupation and average current using the relations of Ref.~\cite{Sankar2025Direct}: 
\begin{equation}
\alpha_{est}=\frac{V_d}{4 \langle N_d\rangle (1- \langle N_d \rangle)}(\frac{\partial \langle N_d\rangle }{\partial \ed})
\label{eq:aocfromocc}
\end{equation}
and
\begin{equation}
\alpha_{est}=V_d(\frac{\partial ln(I_d) }{\partial V_d})
\label{eq:aocfromcond}
\end{equation}
both evaluated the centre of the bias window, $\ed=(\mu_L+\mu_R)/2$. Here \ed follows the gate-voltage-increasing convention of Sec. S1.3.2; Eq.~\ref{eq:aocfromocc} therefore carries the opposite overall sign to Eq. (17) of Ref.~\cite{Sankar2025Direct}, which is written in terms of the bare \ed.

For the experimentally-determined $\alpha_{est}$ (open markers), this calculation is made using experimental data. In order to make a direct comparison to theory that accounts for experimental parameters (finite $T$ and large $V_d$), average occupation and average current are approximated numerically using the framework of Ref.~\cite{Sankar2025Direct}, including experimental $T$ and $V_d$ in addition to the device parameters.  Then $\alpha_{est}$ is calculated as described above from these numerically approximated quantities.

The subsections below describe how the derivatives used to calculate $\alpha_{est}$ are extracted from the data.

\subsection{From $\langle N_d\rangle $}
For the time-resolved data of Fig.~\ref{fig:tr}, the window centre is located empirically for each \eq by computing the midpoint of the gate steps at which the occupation trace crosses 0.2 and 0.8, without relying on the absolute \ed calibration. The derivative in Eq.~\ref{eq:aocfromocc} then comes from a local polynomial fit about this point: locations far from $\eq=0$, where AOC is weak and the trace stays close to the sharp weak-coupling lineshape, are fit with a straight line; locations near $\eq=0$, where AOC visibly rounds the transition, are fit with a cubic polynomial over a correspondinly wider window, so the curvature is captured rather than averaged over. The resulting slopes give the markers in the inset of Fig.~\ref{fig:tr}d. For the average-current occupation maps (Figs.~\ref{fig:scd},\ref{fig:4}), the same window-centre procedure is used, but with a single linear fit over a small, fixed window (11 gate points) at every \eq.
\subsection{From $I_d$}
Every $I_d(V_d)$ trace entering this analysis is itself an average of 150 repeated bias sweeps at fixed \ed, \eq; no additional alignment is needed, since $V_d$, unlike a gate voltage, does not drift between repeats. For the traces of Fig.~\ref{fig:4}c, taken at fixed $\ed=(\mu_L+\mu_R)/2$ while $V_d$ is itself swept, Eq.~\ref{eq:aocfromcond} is evaluated by a local linear fit to $ln(I_d)$ about a particular bias voltage; Fig.~\ref{fig:4}d reports this at $V_d=30$ and $50 \ \mu $V. A complementary view of the same quantity comes from smoothing $ln(\abs{I_d})$ and numerically differentiating, without any local fit, over the full (\eq,$V_d$) plane of Fig.~\ref{fig:diamonds}a; the result is shown in Fig.~\ref{fig:alphas}d. Because this map includes regions where the sequential-tunnelling power law does not strictly apply (such as near $V_d=0$, and deep in Coulomb Blockade), it illustrates the overall trend rather than a quantitative estimate for $\alpha$ at every point; the reported values in the main text and Fig.~\ref{fig:4}d come only from the local-fit procedure above. 

\section{AOC-modified Transport Signatures in Coulomb Blockade Spectroscopy}
\begin{figure*}[h]
    \includegraphics[width=\textwidth]{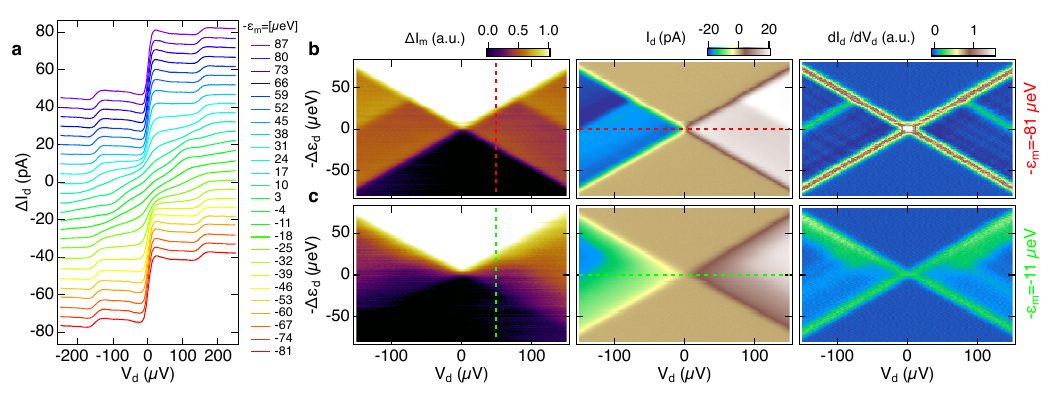}
    \caption{\textbf{Coulomb Blockade Spectroscopy} \textbf{a}, Waterfall of $I_d(V_d)$ for various detector detunings \eq, highlighting the full spectrum over the Coulomb resonance. \textbf{b}, \textbf{c}, Simultaneous Coulomb diamonds in normalized detector current $\Delta I_{m}$, transimission through the dot $I_{d}$, and tunnelling spectroscopy $\frac{dI_{d}}{dV_d}$, for a detector in the weak and strong AOC regime, respectively.}
    \label{fig:diamonds}
    \end{figure*}

Figure~\ref{fig:diamonds} surveys the AOC modifications to transmission $I_d$, shown in Fig.~\ref{fig:4}, in more detail.
Panel a shows the full $I_d(V_d)$ waterfall across 25 values of \eq spanning the detector resonance. This dataset (Fig.~\ref{fig:diamonds}a) is also the source of the $\alpha$ map in Fig.~\ref{fig:alphas}d.

Two \eq settings close to the strong and weak AOC limits shown the main text, $-\eq=-81 \ \mu e$V (weak AOC) and $-\eq=-11 \ \mu e$V (strong AOC), are expanded as complete  Coulomb diamonds in panels b and c. There, the same two settings are compared directly in terms of three simultaneously-measured quantities: normalized detector current $\Delta I_m$, dot current $I_d$, and its derivative $\frac{dI_{d}}{dV_d}$. At weak AOC (upper row), the diamonds remain sharp for all three quantities. This is consistent with the flat, thermally-limited plateaus of Fig.~\ref{fig:4}a. At strong AOC (lower row), the same edges are visibly rounded, most clearly in the tunnelling spectroscopy $\frac{dI_{d}}{dV_d}$. The same rounding used to extract $\alpha$ estimates in Sec. S4, now shown over the full (\ed,$V_d$) plane rather than at one fixed detuning, gives a picture of the dissipation into detector leads induced by the mechanism of AOC-supported inelastic processes. 

\newpage
\section{Robustness of AOC Signatures}
 \subsection{Occupation Features}
\begin{figure*}[h]
    \includegraphics[width=\textwidth]{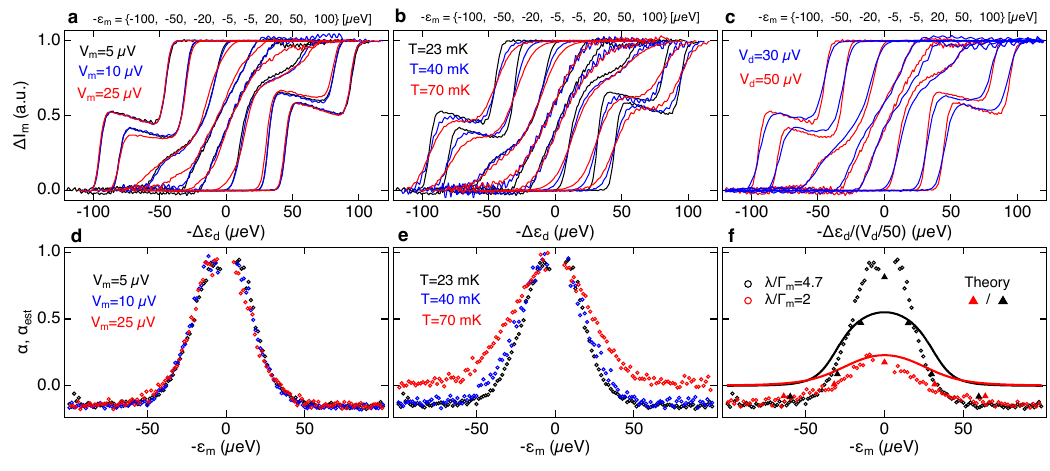}
     \caption{\textbf{Effect of finite temperature and bias voltages.} \textbf{a}, Dependence of normalized average detector current as a proxy for $\langle N_d \rangle$ in terms of detector bias $V_m$, for various detector detunings \eq. \textbf{b}, Dependence of estimated $\langle N_d \rangle(\ed,\eq)$ on varying $T$. The temperatures quoted above base temperature correspond to the mixing chamber temperature. \textbf{c}, $\langle N_d \rangle(\ed,\eq)$ for two different values $V_d$, for which the AOC exponents are shown in Fig.~\ref{fig:4}d. All traces corresponds to the regime shown in Fig.~\ref{fig:scd}a. \textbf{d, e}, Extracted $\alpha_{est}$ from charge susceptibility measurements in panels \textbf{a} and \textbf{b}, highlighting the roles of finite $V_m, T$. \textbf{f}, Comparison of $\alpha_{est}$ for varying $\lambda/\Gamma_m$, corresponding to the settings of Fig.~\ref{fig:scd}a and Fig.~\ref{fig:scd}d, highlighting the overall weaker AOC from a less sensitive detector in Fig.~\ref{fig:scd}d. Expected $\alpha$ as determined from Eq.~\ref{eq:AOC_delta} using energy scales extracted from the charge stability diagrams (solid lines), and $\alpha_{est}$ from numerical estimates (triangle markers) are shown for comparison.}
    \label{fig:params}
    \end{figure*}

Ref.~\cite{Sankar2025Direct} derives the relations Eq.~\ref{eq:aocfromocc} and \ref{eq:aocfromcond} to extract $\alpha$ signatures from data at $k_BT=eV_{m}=0$, showing analytically that they remain reliable in the regime $\{k_BT,eV_m\} \ll eV_d < \Gamma_m$.  However, when $V_d$ rises above the detector bandwidth $\Gamma_m$, and/or when operating at finite $k_BT$ that is not $\ll eV_d$, deviations are expected that would make the values $\alpha_{est}$ extracted from Eq.~\ref{eq:aocfromocc} and \ref{eq:aocfromcond} deviate from the real $\alpha$.   Figures~\ref{fig:tr}d (inset) and Fig.~\ref{fig:4}d compare the expected real $\alpha$ (based on device parameters using Eq.~\ref{eq:AOC_delta}), the $\alpha_{est}$ values extracted from numerical approximations of $\langle N_d \rangle(\ed,\eq)$ and $I_d (\ed,\eq)$ using Eq.~\ref{eq:aocfromocc} and \ref{eq:aocfromcond} but taking experimental $V_d$ and $T$ into account, and the values extracted from the data using Eq.~\ref{eq:aocfromocc} and \ref{eq:aocfromcond}. Further experimental data exploring the degree to which $T$, $V_m$ and $V_d$ affect the extracted $\alpha_{est}$ are shown in Figs.~\ref{fig:params} and \ref{fig:alphas}.

Figure~\ref{fig:params} shows $\langle N_d \rangle(\ed,\eq)$ measured with the same parameters used for Figs.~\ref{fig:scd}a,b. Panel \ref{fig:params}a shows the effect of detector bias $V_m$, which must be nonzero to read out the dot charge at all but is assumed in the theory to be vanishingly small in order for the detector to be near equilibrium. For all the data in the main text, $V_m$ was maintained at 5~$\mu$V.  $\langle N_d \rangle(\ed,\eq)$ is unchanged within experimental resolution upon increasing $V_d$ to $10 \ \mu $V, demonstrating the 5~$\mu$V setting used for the main text data is in the near-equilibrium limit for the detector. However, at $V_m=25 \ \mu $V  the presence of detector bias begins to introduce additional features in the lineshapes due to shot-noise-induced dephasing.

Panel b investigates the role of temperature, showing how occupation traces change as the electron temperature is raised. At 40 and 70~mK, broadening is apparent at the tails of the bias window.  The slope in the middle of the bias window is effectively unchanged at $T=40 \ $mK, but by $70 \ $mK deviations are apparent.  The implication is that one must operate the device at $k_B T < 0.1eV_d$ in order for temperature not to affect the result.

Panel c compares $\langle N_d \rangle(\ed,\eq)$ for the $V_d=30$ and $50 \ \mu $V cases shown in Fig.~\ref{fig:4}d. The horizontal axis of the graph is scaled by $V_d$ in order to make the width of the bias windows equivalent.  Further work may provide a more fundamental insight into the role of large $V_d$ in affecting the $\langle N_d \rangle(\ed,\eq)$ lineshape, but we note that the data matches the numerical theory very closely in Figs.~\ref{fig:tr}d and \ref{fig:4}d.

\subsection{Extraction of the AOC exponent for varying $V_d$ and $\eq$}
Figure~\ref{fig:params}d,e extend the observed role of detector bias and temperature directly to the signatures of $\alpha_{est}$. Here we see the robustness of $\alpha_{est}$ against finite detector bias, as well as the effect of raising the temperature at 70~mK but no effect at 40~mK.  This serves to verify the equilibrium limit for detector bias, and that the reported $\alpha_{est}$ are in the low bias limit the bias window used here.

Figure~\ref{fig:params}f provides a more direct quantitative test of Eq.~\ref{eq:AOC_delta}, showing how $\alpha_{est}$ changes when $\Gamma_m$ changes from the settings of Fig.~\ref{fig:scd}a to those of Fig.~\ref{fig:scd}d.  The $\lambda/\Gamma_m=4.7$ data and theory (black) are the same as shown in Fig.~\ref{fig:4}d, whereas the $\lambda/\Gamma_m=2$ data and theory (red)are the equivalent for higher $\Gamma_m$.  As in Fig.~\ref{fig:4}d, the data deviates from the exact prediction of Eq.~\ref{eq:AOC_delta} but is comparable to the numerical estimates accounting for experimental conditions ($k_BT<eV_d$, $eV_d>\Gamma_m$)

\begin{figure*}[h]
    \includegraphics[width=\textwidth]{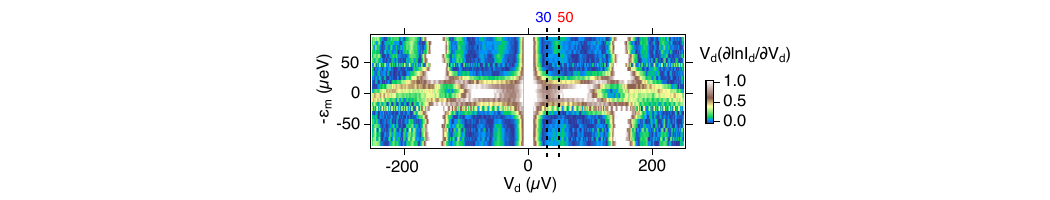}
    \caption{\textbf{Extended calculation of Eq.~\ref{eq:aocfromcond}.} Computed $V_d(\frac{\partial ln(I_d) }{\partial V_d})$ for the full $I_d(V_d)$ spectrum shown in Fig.~\ref{fig:diamonds}a.}
    \label{fig:alphas}
    \end{figure*}

Finally, we present a more exploratory picture in Figure~\ref{fig:alphas}, where Eq.\ref{eq:aocfromcond} is calculated over the full range of $I_d(V_d)$ traces shown in Fig.~\ref{fig:diamonds}a. We emphasize that the validity of this equation in determining $\alpha$ is limited to the low $V_d$ regime, but this characterization of the data may be helpful in identifying phenomena for future investigation.  Dashed lines at 30 and 50 $\mu$V indicate the positions at which the Fig.~\ref{fig:4}d AOC exponents were extracted. Features at large $V_d$, in particular the evolution of the spinful excited state for \eq near zero, is left to future work. 

\renewcommand{\figurename}{Extended Data Fig.}
\renewcommand{\thesubsection}{S\arabic{subsection}}
\setcounter{secnumdepth}{2}
\setcounter{figure}{0} 
\setcounter{equation}{0}

\end{document}